\documentclass[twocolumn,aps,prd,amssymb,eqsecnum,nofootinbib]{revtex4}  
\usepackage{amsmath}
\usepackage{amssymb}
\begin{document}
\title{Simple sufficient conditions for the generalized covariant
  entropy bound}  
\author{Raphael Bousso$^{1,2,3}$, \'Eanna \'E. Flanagan$^{2,4}$ and
  Donald Marolf$^5$} 
\affiliation{$^1$Harvard University,
Department of Physics, Jefferson Laboratory,
17 Oxford Street, Cambridge, MA 02138}
\affiliation{$^2$Radcliffe Institute for Advanced Study, Putnam House,
10 Garden Street, Cambridge, MA 02138}
\affiliation{$^3$On leave from the University of California at Berkeley}
\affiliation{$^4$Cornell University, 
Newman Laboratory, Ithaca, NY 14853-5001}
\affiliation{$^5$Physics Department, Syracuse University, Syracuse, NY13244}

\begin{abstract}

The generalized covariant entropy bound is the conjecture that for any
null hypersurface which is generated by geodesics with non-positive
expansion starting from a spacelike 2-surface $B$ and ending in a
spacelike 2-surface $B^\prime$, the matter entropy on that
hypersurface will not exceed one quarter of the difference in areas,
in Planck units, of the two spacelike 2-surfaces.  We show that this
bound can be derived from the following phenomenological assumptions:
(i) matter entropy can be described in terms of an entropy current
$s_a$; (ii) the gradient of the entropy current is bounded by the
energy density, in the sense that $|k^a k^b \nabla_a s_b| \le 2 \pi
T_{ab} k^a k^b / \hbar$ for any null vector $k^a$ where $T_{ab}$ is
the stress energy tensor; and (iii) the entropy current $s_a$ vanishes
on the initial 2-surface $B$.  We also show that the generalized
Bekenstein bound---the conjecture that the entropy of a weakly
gravitating isolated matter system will not exceed a constant times
the product of its mass and its width---can be derived from our
assumptions.  Though we note that any local description of entropy has
intrinsic limitations, we argue that our assumptions apply in a wide
regime.  We closely follow the framework of an earlier derivation, but our
assumptions take a simpler form, making their validity more transparent in
some examples.

\end{abstract}


\maketitle

\def\beq{\begin{equation}}
\def\endeq{\end{equation}}

\section{Introduction and Summary}

\label{sec:intro}

The covariant entropy bound~\cite{Bou99b} is a conjecture relating the
area of two-dimensional spacelike surfaces to the entropy content of
adjacent regions.  Non-expanding light rays emanating orthogonally
from a spacelike 2-surface $B$ of area $A$ generate a null
hypersurface $L$ called a light sheet.  [The light rays must be
terminated before they begin to expand; this typically occurs when
neighboring rays intersect.]  The covariant bound is the conjecture
that the matter entropy $S_L$ on any such light sheet $L$ satisfies
\footnote{The conjecture requires that semiclassical
corrections to general relativity are negligible in the construction
of $L$.  Strominger and Thompson~\cite{StrTho03} have recently
proposed a modification to the covariant bound and conjectured that it
accommodates one-loop quantum corrections to the geometry.  Using
suitable adaptations of the assumptions and of the proof given here,
this modified bound can be derived in the context of the CGHS model\
\cite{StrTho03}.}
\beq   
S_L\le\frac{A}{4G_{\rm N}\hbar},
\label{ceb}
\endeq
where $G_{\rm N}$ is Newton's constant\footnote{We use units with $c =
k_{\rm B} =1$, but we retain Newton's constant $G_{\rm N}$ and
Planck's constant $\hbar$.  We use the $(-,+,+,+)$ metric signature.
Also we work in four spacetime dimensions, although our analysis
generalizes easily to $D$ spacetime dimensions for all $D \ge 3$.}.

When gravity is weak, this bound reduces to some earlier, more
specialized proposals~\cite{Tho93,Sus95}, which originally arose in
attempts to allow the preservation of quantum-mechanical unitarity in
the presence of black holes.  It also generalizes and refines a
cosmological bound conjectured by Fischler and
Susskind~\cite{FisSus98}.  No general derivation of the covariant
bound (or of the earlier bounds) has been given.  However, no
realistic counterexamples have been constructed from known matter
fields.  Moreover, the bound has been verified explicitly in a variety
of examples, including both weakly gravitating and gravitationally
collapsing thermodynamic systems, as well as cosmological spacetimes.
In addition, sufficient conditions for the bound have been identified
which are readily seen to hold in a wide class of
situations~\cite{FMW}.

From a conceptual point of view, there is a tension between the
scaling of maximal entropy with area asserted by the conjecture and
the extensivity of quantum field theory.  The holographic principle
proposes that quantum gravity contains features that resolve this
tension and that give rise to the covariant bound.  Specifically, the
holographic principle asserts that in the fundamental theory, all
physics on light sheets that have maximal area $A$ can be described by
roughly $A$ binary degrees of freedom (in Planck
units)~\cite{Tho93,Sus95,Bou99c}. A detailed review of the covariant
entropy bound and related issues can be found in Ref.~\cite{Bou02}.

In Ref.~\cite{FMW} a stronger form of the covariant bound was
suggested, which we call the generalized covariant entropy bound
(GCEB).  Consider a light sheet $L$ some of whose generators are
terminated prematurely (before they reach caustics).  The endpoints of
these prematurely-terminating generators form a second spacelike
2-surface $B^\prime$ with nonvanishing area $A'$.  The GCEB is the
conjecture that the matter entropy $S_L$ on such a light sheet
satisfies
\begin{equation}
S_L \le \frac{A- A^\prime}{4G_{\rm N}\hbar}.
\label{gceb}
\end{equation}
This bound reduces to the original covariant bound in the special case
$A^\prime=0$.  A key motivation for considering this generalization of
the covariant entropy bound is that the GCEB, if true, would imply as
a special case the generalized second law in classical regimes where
the null energy condition is satisfied~\cite{FMW}.  

Moreover, it has recently been shown~\cite{Bou03} that the GCEB
implies a generalized version of Bekenstein's bound~\cite{Bek74,Bek81}
on the entropy $S$ of any weakly gravitating isolated matter system of
mass $M$ traversed by a light sheet $L$ of initially vanishing
expansion:
\begin{equation}
S \leq \pi M x/\hbar.
\label{eq-gbek}
\end{equation}
The width $x$ is defined as the longest distance (in the center of
mass frame) traveled by any of the generators of $L$ between entering
and exiting the system.  In the context of the original Bekenstein
bound, one usually considers static, compact matter systems.  For such
systems, $x$ is always smaller than the diameter $d$ of the smallest
sphere circumscribing the system.  Moreover, if the GCEB holds for
each light sheet traversing such a system along every spatial
direction, $x$ can be minimized by judicious orientation of the system
relative to the light sheet.  Therefore the bound (\ref{eq-gbek}) is
stronger than the original bound $S \leq \pi M d/\hbar$ advocated by
Bekenstein
\cite{Bek74}.

Since the GCEB also implies the original covariant bound (\ref{ceb}),
it is the strongest of the various conjectured bounds we have just
reviewed.  If entropy bounds are indeed related to a fundamental
theory, this primacy would suggest that the GCEB bears one of that
theory's more direct imprints.  In any case, it is important to
investigate whether or not nature obeys the GCEB.

The purpose of this paper is to establish the validity of the GCEB in
a broad class of hydrodynamic regimes, that is, regimes in which
matter admits a description in terms of continuum variables.  [Note
that we do not exclude solids.]  Specifically, we consider only
regimes in which matter entropy can be described in terms of a local
entropy current $s^a$.\footnote{We will use the term ``current'' for a
flow in the time direction.  For typical matter, $s^a$ will be a
future-directed vector field.  For example, in the local comoving
frame of a fluid, the time component of $s^a$ is the usual entropy
density, and the spatial components vanish.} In such regimes, any
appropriate fundamental definition of the total matter entropy $S_L$
on a light sheet $L$ should, to a good approximation, reduce to the
integral
\begin{equation}
S_L = \int_L s^a \epsilon_{abcd}
\label{eq-sl0}
\end{equation}
of the entropy current over $L$ \footnote{Here the orientation chosen
for $L$ is that which gives a positive result for future-directed,
timelike $s^a$.}.  With entropy given by this formula, we will show
that the GCEB can be derived from two postulated properties of the
entropy current [Eqs.\ (\ref{assumption1}) and (\ref{assumption2})
below].  We will also argue that those assumptions should be valid in
a large class of hydrodynamic regimes.  Our approach is similar in
spirit to that of Ref.~\cite{FMW}. However, the assumptions used here
are somewhat simpler than the assumption that was used in
Ref.~\cite{FMW} to derive the GCEB.

Before attempting to prove the GCEB in hydrodynamic regimes, we must
confront a crucial limitation of the entropy current description.  In
Sec.~\ref{sec:guedens} we note that apparent violations of the GCEB
can be obtained by integrating any finite entropy current over
sufficiently small distances.  However, a hydrodynamic description of
matter entropy becomes invalid at sufficiently small scales.  We
analyze a cosmological example due to Guedens~\cite{Gue00} in which
the onset of apparent violations of the GCEB is seen to coincide with
the ultraviolet breakdown of the hydrodynamic description.

The analysis of Sec.\ \ref{sec:guedens} informs our choice of
assumptions concerning the behavior of the entropy current $s^a$,
which we present in Sec.~\ref{assumptions}.  We argue that those
assumptions should be valid in a large class of hydrodynamic regimes.
Moreover, we note that the assumptions effectively impose an
ultraviolet cutoff related to the local energy density.  This
effective cutoff allows our proof to evade the generic short-distance
problems of the hydrodynamic approximation.  In Sec.~\ref{sec-gceb} we
derive the GCEB from our assumptions.  Finally, in
Sec.~\ref{bekenstein}, we give a direct, purely non-gravitational
derivation of the generalized Bekenstein bound (\ref{eq-gbek}) from
the same assumptions, without using the GCEB as an intermediate
result.

We stress that this paper has nothing to say about the validity of the
GCEB or of the generalized Bekenstein bound in regimes where our
assumptions are not satisfied.  Moreover, neither Ref.~\cite{Bou02}
(which demonstrated that the GCEB implies Bekenstein's bound) nor the
present paper (which shows that both bounds can be derived from
plausible assumptions in a hydrodynamic regime) bear on the
much-debated question of whether any of the proposed entropy bounds
follow from the generalized second law of
thermodynamics~\cite{Wal99,Bek01,MarSor02,MarSor03}.

Our result that the GCEB is valid in hydrodynamic regimes, together
with the results of Ref.~\cite{FMW}, eliminates a large set of
possible counterexamples to the GCEB.  However, if entropy bounds
really have the significance ascribed to them by the holographic
principle, one would expect the GCEB (or at least the original
covariant bound) to apply more broadly.  If it does, a complete proof
may not be possible until the underlying quantum gravity theory is
understood.  In the meantime, further exploration of the bounds'
domain of validity remains an important task which may produce clues
about quantum gravity.

\section{Intrinsic limitations of a local description of entropy}
\label{sec:guedens}

On sufficiently large scales a description of entropy can often be
given in terms of a local entropy current $s^a$.  In such situations,
the actual entropy contained in any spatial region ${\cal R}$ is, to a
good approximation, given by the integral of the entropy current over
${\cal R}$, as long as ${\cal R}$ is much larger than some
microphysical length scale $\Lambda$ which is determined by the
physical system under consideration.  For example, for a bath of
thermal radiation, $\Lambda$ is of order the wavelength of a typical
quantum of radiation.  Thus, the entropy current $s^a$ makes sense
only when integrated over sufficiently large regions.  It is important
not to take this local description of entropy too literally.  Entropy
is fundamentally nonlocal, and there is no physical justification for
computing an entropy for regions smaller than $\Lambda$ by integrating
the entropy current $s^a$.

Indeed, if the entropy-current approximation for entropy could be
extrapolated to arbitrarily short distances, the GCEB could clearly be
violated.  Consider an initial 2-surface $B$ on which the expansion
$\theta$ vanishes everywhere.  Construct a very short light sheet $L$, for
which the final 2-surface $B^\prime$ is allowed to approach the
initial 2-surface $B$ arbitrarily closely, so that the affine
parameter interval $\Delta\lambda$ along the null generators goes to
zero.  Then the change $A - A^\prime$ in area is quadratic in $\Delta
\lambda$, since the area of cross sections of $L$ has a local maximum
at $B$.  But in the entropy-current approximation, the total entropy
$S_L$ is given by the integral (\ref{eq-sl0}) and scales linearly with
$\Delta \lambda$ for small $\Delta\lambda$:
\begin{equation}
S_L \propto \Delta\lambda.
\label{eq-sdl}
\end{equation}
Therefore the GCEB will be violated for sufficiently small
$\Delta\lambda$.  It follows in particular that the GCEB cannot be
derived from any assumptions that permit the integration of a
nonvanishing entropy current over arbitrarily short light sheets with
initially vanishing expansion.

An example of this type was first pointed out by Raf
Guedens~\cite{Gue00}.  Its original purpose was to show that the GCEB
cannot be derived from the second set of assumptions (1.10) and (1.11)
of Ref.~\cite{FMW}, as the example explicitly satisfies those
assumptions.  It is instructive to review the Guedens example before
attempting to formulate more suitable assumptions.

Consider a closed, radiation-dominated Friedman-Robertson-Walker
cosmological model, for which the metric is
\begin{equation}
ds^2 = a(\eta)^2 \left[ - d\eta^2 + d \chi^2 + \sin^2 \chi
 ( d\vartheta^2
  + \sin^2 \vartheta d\varphi^2) \right],
\end{equation}
with scale factor
\begin{equation}
a(\eta) = a_{\rm m} \sin \eta.
\end{equation}
Here $a_{\rm m}$ is the radius of the spatial 3-sphere at the the
moment $\eta = \pi/2$ of maximum expansion.  Next, we choose a light
sheet $L$ and compute its area decrease, $A-A'$, and entropy, $S_L$.
Let $L$ begin on the 2-sphere $B$ of maximum radius in the spacetime,
given by $\chi =\eta =\pi/2$, and let it end on the nearby 2-sphere
$B^\prime$ given by $\chi =\eta =\pi/2 +\Delta\chi$.

We shall work to leading order in $\Delta \chi$, which means that we
can use $\Delta \chi$ as an affine parameter.  The area decrease is
quadratic:
\begin{equation}
A - A^\prime = 8 \pi a_{\rm m}^2 \Delta \chi^2 [1+O(\Delta\chi)].
\label{guedensarea}
\end{equation}
In the fluid approximation, the total entropy $S_L$ is just the
product of the entropy density\footnote{We use $s$ to denote an
entropy density per unit affine parameter and per unit proper
transverse area [see Eq.~(\ref{eq:densitydef})].  Here it was
convenient to choose the normalization of the affine parameter
$\Delta\chi$ such that $s$ coincides with the usual entropy per unit
proper volume (in the local rest frame of the cosmological fluid).
Note that this choice differs from the normalization convention we
will use in our proof in Sec.\
\protect{\ref{sec:proof}} below.} $s$ and the volume $4 \pi
a_{\rm m}^3\Delta\chi$ of the projection of $L$ onto the $\eta =
\pi/2$ hypersurface.  For a radiation-dominated universe, standard
thermodynamics implies
\begin{equation}
s = \frac{4 \rho}{3 T},
\label{edens}
\end{equation}
where $\rho=3 / (8 \pi G_{\rm N} a^2)$ is the energy density and $T$
the temperature.  This yields
\begin{equation}
S_L = {2 a_{\rm m} \Delta \chi \over G_{\rm N} T} [1+O(\Delta\chi)].
\label{guedensentropy}
\end{equation}
Comparison with the area change (\ref{guedensarea}) shows that the
bound (\ref{gceb}) is apparently violated when
\begin{equation}
a_{\rm m} \Delta \chi \le {\hbar \over \pi T}.
\label{guedenscondition}
\end{equation}

Equation (\ref{guedenscondition}) says that the proper length $a_{\rm
m} \Delta \chi$ in the cosmological rest frame of the light sheet is
shorter than the thermal wavelength $\hbar/T$.  Therefore the light
sheet is shorter than even a single wavelength of the radiation
filling the Universe.  In this regime it is clear that our computation
of the entropy (\ref{guedensentropy}) is invalid.

Thus, the onset of apparent violations of the GCEB coincides in the
Guedens example with the breakdown of the local description of
entropy, namely when the condition (\ref{guedenscondition}) holds.  In
the regime (\ref{guedenscondition}), most of the occupied modes cannot
be localized within $L$ but will spill over beyond the boundaries of
the light sheet.  Thus we cannot conclude that the GCEB is violated in
any physically meaningful sense.

A number of questions arise concerning the applicability and precise
formulation of the GCEB at short distances, such as the regime
(\ref{guedenscondition}) in the Guedens example.  Is there a
meaningful definition of entropy for such short light sheets?  It is
conceivable that the entropy $S_L$ computed from a sufficiently
general and fundamental prescription will satisfy the GCEB at all
scales.  Alternatively, it is possible that the GCEB will only apply
to the statistical entropy of complete, isolated systems.  If this is
true, one would need to supplement the statement of the GCEB with a
suitable criterion defining ``isolation''.  Resolution of these issues
would sharpen the conjecture and may contribute to a deeper
understanding of its meaning.

\section{Derivation of entropy bounds}
\label{sec:proof}

In this section we will derive the GCEB and the generalized Bekenstein
bound in the regime defined by Eq.~(\ref{eq-sl0}), subject to two
assumptions.  We start by introducing some notations to describe the
integral (\ref{eq-sl0}) over the light sheet $L$.  The affine
parameter, $\lambda$, along any null generator of $L$ is taken to run
from $0$ at $B$ to $1$ at $B'$ \footnote{We neglect any generators of
$L$ whose affine parameter length is infinite; this is justified in
Ref.\ \protect{\cite{FMW}}.}.  The null vector $k^a\equiv
(d/d\lambda)^a$ is normal (and tangent) to $L$.  We introduce a
coordinate system $x = (x^1,x^2)$ on the initial 2-surface $B$, and we
label the geodesic generators of $L$ with these coordinates, thereby
defining a coordinate system $(x^1,x^2,\lambda)$ on $L$.  We denote by
$h_{AB}(x,\lambda)$ the two dimensional induced metric on the
cross-sections $\lambda = $ const of $L$, and define $h = {\rm det}\,
h_{AB}$.  With these notations the integral (\ref{eq-sl0}) can be
written as \cite{FMW}
\begin{equation}
S_L = \int_B d^2
x \, \int_0^1 d\lambda \, \sqrt{h(x,\lambda)} s(x,\lambda).
\label{eq-sl}
\end{equation}
Here $s$ is the the entropy density on $L$, or more precisely the
entropy per unit affine parameter and per unit cross-sectional
area.  It is given by  
\begin{equation}
s = \pm k^a s_a,
\label{eq:densitydef}
\end{equation}
where the minus sign applies for future directed $k^a$ and the plus sign
for past directed $k^a$.
The determinant factor in Eq.\ (\ref{eq-sl}) can be written as \cite{FMW}
\begin{equation}
\sqrt{h(x,\lambda)} = {\cal A}(x,\lambda)\sqrt{h(x,0)},
\end{equation}
where ${\cal A}$ is an area decrease factor associated with a given
generator given by
\begin{equation}
{\cal A}(\lambda) \equiv \exp \left[ \int_0^\lambda d{\bar \lambda}\,
  \theta({\bar \lambda}) \right].
\label{eq:calAdef}
\end{equation}
Here $\theta = \nabla_a k^a$ is the expansion of the generators of $L$
which by assumption is nonpositive everywhere on $L$.  A prime will be
used to denote the operator $k^a \nabla_a = d/d\lambda$.

\subsection{Assumptions}
\label{assumptions}

We assume that the entropy density $s$ on a light sheet $L$ satisfies
two conditions:
\begin{enumerate}
\item The {\it gradient assumption} that 
\begin{equation}
s' \le 2 \pi T_{ab} k^a k^b/\hbar
\label{assumption1}
\end{equation}
at every point on $L$, where $T_{ab}$ is the stress tensor.
\item The {\it isolation assumption}\footnote{Strominger and Thompson
    have shown that this assumption can be replaced with the
    requirement that $s \le -\theta/4$ on the initial 2-surface, which
    can be interpreted as demanding that the GCEB is satisfied in an
    infinitesimal neighborhood of $B$ \protect{\cite{StrTho03}}.} that
    the entropy density should vanish on the initial 2-surface:
\begin{equation}
s_{|B} =0.
\label{assumption2}
\end{equation}
\end{enumerate}

We now discuss the motivation for these assumptions.  The first
assumption essentially states that the entropy gradient must be
smaller than the energy density in natural units.  This
holds for free Bose and Fermi gases in local thermal
equilibrium~\cite{FMW}.  More generally, the assumption is plausible
if (i) both the entropy density and the stress tensor are smeared (in
all spatial directions) over a distance $\Lambda$ set by the largest
wavelength of any of the modes that contribute significantly to the
entropy, and if (ii) the effective number of scalar fields in the
Lagrangian, $N$, is not very large.  In this case, one can apply an
estimate given in Ref.~\cite{FMW} which we refine here.

We neglect factors of order unity.  Because of the lack of features at
distances smaller than $\Lambda$, the entropy gradient obeys $\nabla
s\lesssim s/\Lambda$.  Now consider a sphere of radius $\Lambda$.
Since $\Lambda$ is the largest wavelength, particles can be considered
localized on this scale, and we can define $n$ to be the number of
particles inside the sphere.  Since we assume that modes with
wavelength $\Lambda$ contribute significantly to the entropy, we can
obtain the order of magnitude of the entropy by considering only such
modes.  Then the number of states will be $(N+n-1)!/[(N-1)! n!]$ [or
$(N+n)!/(N! n!)$ if states with fewer than $n$ particles are also
allowed].  The total entropy ${\cal S}$ in the sphere is the logarithm
of the this number and hence obeys ${\cal S}\leq n \ln N$ (saturation
occurs for small $n$).  Unless the number of fields is very large,
$\ln N$ will be of order unity.  The entropy density therefore
satisfies $s\lesssim n/\Lambda^3$.  On the other hand, the energy
density is bounded from below: $\rho\gtrsim n\hbar/\Lambda^4$.  It
follows that $\nabla s\lesssim\rho/\hbar$.

The gradient assumption is clearly closely related to Bekenstein's
bound.  One might even interpret it as a kind of local formulation of
Bekenstein's bound, or at least as a first step in that 
direction.  The same may be said of the quasi-local assumption (1.9)
of Ref.\ \cite{FMW}, which was shown in that reference to
independently imply the GCEB.

Consider now the second assumption (\ref{assumption2}).  It is clearly
satisfied if the initial 2-surface $B$ is in a vacuum region outside a
compact matter system.  If the initial 2-surface lies instead inside
the matter system so that the initial entropy current $s_{a\,|B}$ is
nonvanishing, then the assumption is violated.  However, we can
imagine a slightly different matter system in which the the entropy
current in a thin sliver near the initial 2-surface $B$ has been
modified to achieve $s_{a\,|B}=0$ without violating the gradient
assumption (\ref{assumption1}).  This will only require changing the
entropy current within one or two wavelengths of the initial surface,
and thus the change in the total integrated entropy will be negligible
(so long as the extent of $L$ is significantly larger than one
wavelength).  But the fluid approximation is in any case valid only
for regions much larger than a typical wavelength.  Hence, our second
assumption poses no significant additional restriction on the range of
applicability of our proof.

In the previous section we pointed out that the hydrodynamic
description of entropy breaks down at short distances, where it would
generically lead to violations of the GCEB.  These difficulties will
not plague our derivations, as our two assumptions together
exclude the arbitrarily short light sheets with finite entropy density
that led to the problematic scaling (\ref{eq-sdl}).  Namely,
Eq.~(\ref{assumption2}) ensures that the entropy density vanishes on
the initial surface, and Eq.~(\ref{assumption1}) prevents it from
turning on too rapidly.

In this sense the second assumption (\ref{assumption2}) addresses the
uncertainty as to the formulation of the GCEB at short distances.  It
can be interpreted to do so via either of the two approaches to this
formulation outlined at the end of Sec.~\ref{sec:guedens}.  Most
straightforwardly, (\ref{assumption2}) may be regarded as expressing
the requirement that the matter on the light sheet $L$ be isolated, in
the sense that modes carrying significant amounts of entropy should be
fully contained on $L$ and should not spill over beyond the initial
boundary of $L$.

The second interpretation is to regard (\ref{assumption2}) as an
imprint of a more general prescription for defining the entropy $S_L$
through $L$.  By insisting on setting $s=0$ on the initial boundary
$B$, and allowing $s$ to increase only at a rate limited by
Eq.~(\ref{assumption1}), the contributions of ``spill-over'' modes are
effectively removed from the vicinity of the boundary.  Note that
under such a prescription, $s$ could not be obtained as the contraction
$k^a s_a$ of $k^a$ with an absolute entropy current $s^a$ that is the
same for all light sheets.  Rather, such a
prescription would entail light-sheet dependent entropy
currents.  However, in view of the nonlocal nature of entropy, some
kind of sub-additive, light-sheet dependent prescription may indeed be
appropriate.  Light-sheet dependent currents have previously been
considered, with a similar motivation, in the assumption (1.9) of
Ref.~\cite{FMW}.  Note that if $s$ does arise from a global, absolute
entropy current vector field $s^a$, our second condition
(\ref{assumption2}) will be satisfied for every light sheet in the
spacetime if and only if $s^a$ satisfies $|k^a k^b\nabla_a s_b| \leq
2\pi T_{ab} k^a k^b/\hbar$ for all null vector fields $k^a$ (the
condition cited in the abstract).  We stress, however, that our proof
applies both to light-sheet dependent and absolute entropy currents;
as long as $s$ satisfies the required conditions on some light sheet
$L$, we prove that the GCEB will hold on $L$.

\subsection{Derivation of the generalized covariant bound}
\label{sec-gceb}

In Ref.\ \cite{FMW} it was shown that, to prove the GCEB using a local
entropy current, it is sufficient to focus on each individual null
generator of the light sheet $L$, one at a time.  Specifically, we
need only show that
\begin{equation}
\int_0^1 d\lambda\, s(\lambda) {\cal A}(\lambda) 
\le {1 \over 4} \left[1  - {\cal A}(1) \right]
\label{condt}
\end{equation}
for each null generator of $L$ of finite affine parameter length,
where ${\cal A}(\lambda)$ is the area-decrease factor
(\ref{eq:calAdef}).

The Raychaudhuri equation along the geodesic can be written in the form
\begin{equation}
8 \pi G_{\rm N} T_{ab} k^a k^b+ {\hat \sigma}_{ab} {\hat\sigma}^{ab}
 = {- 2G^{\prime\prime}(\lambda) \over G(\lambda)},
\label{ray}
\end{equation}
where $G(\lambda) \equiv \sqrt{{\cal A}(\lambda)}$ and ${\hat
\sigma}_{ab}$ is the shear tensor, and primes denote derivatives with
respect to $\lambda$.   Thus the gradient assumption
(\ref{assumption1}) implies that\footnote{Note that the corresponding
Eq.\ (2.17) of Ref.\ \protect{\cite{FMW}} contains a typographical
error; the role of the barred and unbarred quantities should be
interchanged.}
\begin{equation}
 s^\prime(\lambda) \le  \frac{-1}{2 G_{\rm N} \hbar} 
  {G^{\prime\prime}(\lambda) \over  G(\lambda)}.
\label{eq4}
\end{equation}
Using our second assumption, $s(0)=0$, we integrate this expression to
obtain a bound on the scalar entropy density:
\begin{eqnarray}
s(\lambda) &=&  \int_0^\lambda d{\bar \lambda}\, s'({\bar \lambda})
\nonumber \\
&\le&  {-1 \over 2 G_{\rm N} \hbar} \int_0^\lambda d{\bar \lambda}\, {
G^{\prime\prime}({\bar \lambda}) \over G({\bar \lambda})}.
\end{eqnarray}
Integration by parts now gives
\begin{equation}
s(\lambda) \le {1 \over 2 G_{\rm N} \hbar} 
\left[ {G^\prime(0) \over G(0)} -
 {G^\prime(\lambda) \over G(\lambda)} - \int_0^\lambda  
 d{\bar \lambda}\, { G^\prime({\bar \lambda})^2 \over G({\bar
 \lambda})^2 }\right].
\label{eq:f}
\end{equation}

The first term in the square brackets in Eq.\ (\ref{eq:f}) is
nonpositive by the non-expansion condition $\theta\leq 0$, since
$G^\prime(0) = \theta(0)/2$.  The last term in the square brackets is
manifestly nonpositive.  Hence both terms 
can be discarded, and we obtain
\begin{equation}
s(\lambda) \le  {-1 \over 2 G_{\rm N} \hbar}
 {G^\prime(\lambda) \over  G(\lambda)}.
\end{equation}
Inserting this into the left hand side of Eq.\ (\ref{condt}) and using
${\cal A} = G^2$ we obtain
\begin{eqnarray}
\int_0^1 d\lambda\, s(\lambda) {\cal A}(\lambda) &\le&
 {-1 \over 2 G_{\rm N} \hbar}\int_0^1 d \lambda\, G(\lambda) 
G^\prime(\lambda) \nonumber \\
\mbox{} &=&  {1 \over 4 G_{\rm N} \hbar}
 \left[ G(0)^2 - G(1)^2 \right].
\end{eqnarray}
Finally using $G(0)=1$ and $G(1)^2 = {\cal A}(1)$ yields Eq.\ 
(\ref{condt}).  This completes the proof.

\subsection{Derivation of the generalized Bekenstein bound}
\label{bekenstein}

We have derived the GCEB for light sheets on which our assumptions
hold, and in Ref.~\cite{Bou02} the GCEB was shown to imply the
generalized Bekenstein bound Eq.~(\ref{eq-gbek}) for weakly
gravitating matter systems.  Curiously, in this sequence of
implications neither our assumptions nor the Bekenstein bound contain
Newton's constant, whereas the GCEB does.  It seems strange that
we should have to go via a bound involving gravity to derive the
Bekenstein bound from our assumptions, using the Raychaudhuri equation
once in Eq.~(\ref{ray}) to introduce $G_{\rm N}$, and then a second  
time in Ref.~\cite{Bou03} to eliminate it.  Here we point
out that it is indeed possible to obtain the generalized Bekenstein
bound directly from our assumptions, without using general relativity.

The derivation of the generalized Bekenstein bound involves two light
sheets $L_\pm$ which nearly coincide with each other but are
oppositely directed~\cite{Bou03}.  Each has initially vanishing
expansion, and each fully contains the matter system.  Specifically,
one requires that the matter system occupies a world volume $W$ of
compact spatial support in approximately Minkowski space and that no
world-line in $W$ fails to intersect $L_\pm$.  As indicated in the
introduction, the width of the system, $x$, is defined in terms of the
light sheets $L_\pm$.  Thus the light sheets $L_\pm$ are an integral
part of of the generalized Bekenstein bound.  [This was not the case
for the original Bekenstein bound, which was formulated mainly for
static systems and employs a different, more lenient definition of
width.]

The GCEB had to be assumed to hold for both $L_+$ and $L_-$ at the
outset of the derivation in Ref.~\cite{Bou03}.  In order to give a
direct, non-gravitational derivation of the generalized Bekenstein
bound in the hydrodynamic regime (\ref{eq-sl0}), we should therefore
assume that our conditions (\ref{assumption1}) and (\ref{assumption2})
hold for $L_+$ and $L_-$.  In Ref.~\cite{Bou03} the assumption of weak
gravity allowed the construction of nearly coinciding light sheets
$L_\pm$ with small relative area decrease.  In the limit $G\to 0$, the
same construction results in exactly coinciding light sheets,
$L_+=L_-=L$ with everywhere vanishing expansion, in flat Minkowski
space.  For a direct non-gravitational derivation of the generalized
Bekenstein bound from our assumptions, we will set $G_{\rm N}$ to zero
from the start and work with the simpler object $L$.

Let $L$ be a congruence of future-directed parallel null geodesics
($\theta=0$) orthogonal to a compact 2-surface $B$.  Let each ray
extend a finite spatial distance (in a fixed Lorentz frame), so that
the congruence ends on a second 2-surface $B'$.  All cross sections of
the congruence including $B$ and $B'$ have the same area $A$, since
spacetime is flat.  Thus, we can regard $L$ as a light sheet $L_+$
originating at $B$, with an affine parameter $\lambda_+$ that varies
from $0$ at $B$ to $1$ at $B^\prime$.  Alternatively, we can regard
$L$ as a light sheet $L_-$ originating at $B^\prime$, with an affine
parameter $\lambda_-$ that varies from $0$ at $B^\prime$ to $1$ at
$B^\prime$.  The two affine parameters are related by
\begin{equation}
\lambda_- = 1 - \lambda_+.
\label{eq:rel}
\end{equation}
We require our two assumptions to hold in either case; in particular,
the isolation condition, $s=0$, is assumed to hold on both terminal
2-surfaces.

The entropy density $s$ at any point on $L$ can thus be computed in
two ways, by integrating $s'$ starting from either end:
\begin{equation}
s(\lambda_+) = \int_0^{\lambda_+} d\bar\lambda_+
\frac{ds}{d\bar\lambda_+}
= \int_0^{1-\lambda_+} d\bar\lambda_- 
\frac{ds}{d\bar\lambda_-} .
\end{equation}
By adding both expressions, applying the gradient assumption to each
integrand, and using $\lambda_-=1-\lambda_+$, we find
\begin{equation}
2s(\lambda_+) \leq 2\pi \int_0^1 d\bar\lambda_+ T_{ab} k^a k^b/\hbar,
\end{equation}
where $k^a = (d/d\lambda_+)^a$.  Since the gradient assumption applies
with respect to both directions, and $ds/d\lambda_+=-ds/d\lambda_-$,
the null energy condition must hold and hence the integrand is
positive definite.  By construction, the direction of $k^a$ is the
same for all light-rays, and we can only increase the right hand side
by replacing $k^b$ with $k^b_{\rm max}$, the null vector with the
largest time component in a fixed Lorentz frame, maximizing over all
the generators of $L$.  Hence,
\begin{equation}
s(\lambda_+) \leq \pi \int_0^1 d\lambda_+ T_{ab} k^a k_{\rm
max}^b/\hbar.
\end{equation}
Further integration over the light sheet $L$ then yields
\begin{equation}
S \leq \pi P_b k_{\rm max}^b/\hbar,
\end{equation}
where $P_b = \int_B \int d\lambda_+ T_{ab} k^a$ is the total
four-momentum of the matter present on $L$~\cite{Bou02}.  This
inequality takes its simplest form in the rest frame of the matter
system, for which the spatial components of $P_b$ vanish.  Then $P_0$
is given by the system's rest mass, $M$, and $k_{\rm max}^0$ is a
proper length corresponding to the width of the system, $x$.  [Recall
that $x$ is defined as the greatest spatial distance traversed by any
of the generators of $L$, in the rest frame of $W$.]  Thus we obtain
the generalized Bekenstein bound (\ref{eq-gbek}).

\acknowledgments
We thank A.~Strominger and D.~Thompson for useful conversations.  This
research was supported in part by the Radcliffe Institute, by NSF
grants PHY00-98747 and PHY-0140209, and by funds from Syracuse
University.

\bibliographystyle{board}
\bibliography{all}

\end{document}